\documentclass[sigplan,nonacm]{acmart}

\def \showcomments {Show comments}
\def \showauthors  {Show authors}
\def \showchanges  {Show changes}

\usepackage{xspace}
\usepackage{balance}
\usepackage[htt]{hyphenat} 
\usepackage[underline=false]{pgf-umlsd} 
\usepackage{subcaption}
\usepackage{cleveref}
\usepackage{enumitem}
\usetikzlibrary{positioning}

\ifdefined \showcomments
\usepackage{todonotes}
\usepackage{letltxmacro}
\LetLtxMacro{\todonote}{\todo}
\renewcommand{\todo}[2][]
{\todonote[inline, caption={#2}, size=\footnotesize, #1]
{\renewcommand{\baselinestretch}{0.5}\selectfont#2\par}}
\else
\usepackage[disable]{todonotes}
\fi

\newcommand{\sys}{CTR\xspace}
\newcommand{\syssgx}{\sys-SGX\xspace}
\newcommand{\Checkpoint}{Checkpoint\xspace}
\newcommand{\checkpoint}{checkpoint\xspace}
\newcommand{\Restore}{Restore\xspace}
\newcommand{\restore}{restore\xspace}

\newcommand{\destination}{destination\xspace}
\newcommand{\ecall}{\texttt{ECALL}\xspace}

\newcommand{\criu}{CRIU\xspace}
\newcommand{\mk}{\texttt{MK}\xspace}

\newcommand{\eg}{e.g.,\xspace}
\newcommand{\ie}{i.e.,\xspace}
\newcommand{\etal}{et al.\xspace}

\ifdefined \showchanges

\else

\fi

\newcommand{\tinyskip}{\vspace{3pt}}
\newcommand{\mypar}[1]{\tinyskip\noindent\textbf{#1.}\xspace}

\makeatletter
\let\@authorsaddresses\@empty
\makeatother

\setcopyright{none}
\renewcommand\footnotetextcopyrightpermission[1]{}

\newlength{\gapspace}

\begin{document}

\title[\sys: Checkpoint, Transfer, and Restore for Secure Enclaves]{\sys: Checkpoint, Transfer, and Restore \\for Secure Enclaves}

\ifdefined \showauthors
\author{Yoshimichi Nakatsuka}
\email{nakatsuy@uci.edu}
\affiliation{%
  \institution{University of California, Irvine}
  \country{USA}
}

\author{Ercan Ozturk}
\email{ercano@uci.edu}
\affiliation{%
  \institution{University of California, Irvine}
  \country{USA}
}

\author{Alex Shamis}
\email{alexsha@microsoft.com}
\affiliation{%
  \institution{Microsoft Research and\\Imperial College London}
  \country{UK}
}

\author{Andrew Paverd}
\email{andrew.paverd@microsoft.com}
\affiliation{%
  \institution{Microsoft Research and\\Microsoft Security Response Center}
  \country{UK}
}

\author{Peter Pietzuch}
\email{prp@imperial.ac.uk}
\affiliation{%
  \institution{Microsoft Research and\\Imperial College London}
  \country{UK}
}
\fi

\begin{abstract}
Hardware-based Trusted Execution Environments (TEEs) are becoming increasingly
prevalent in cloud computing, forming the basis for \emph{confidential
computing}. However, the security goals of TEEs sometimes conflict with
existing cloud functionality, such as VM or process migration, because TEE
memory cannot be read by the hypervisor, OS, or other software on the platform.
Whilst some newer TEE architectures support migration of entire protected VMs,
there is currently no practical solution for migrating individual processes
containing in-process TEEs. The inability to migrate such processes leads
to operational inefficiencies or even data loss if the host platform must be urgently restarted.

We present \emph{\sys}, a software-only design to \emph{retrofit} migration
functionality into existing TEE architectures, whilst maintaining their
expected security guarantees. Our design allows TEEs to be interrupted and
migrated at arbitrary points in their execution, thus maintaining compatibility
with existing VM and process migration techniques. By cooperatively involving
the TEE in the migration process, our design also allows application developers
to specify stateful migration-related policies, such as limiting the number of
times a particular TEE may be migrated. Our prototype implementation for Intel
SGX demonstrates that migration latency increases linearly with the size of the
TEE memory and is dominated by TEE system operations.

\end{abstract}

\settopmatter{printfolios=true}
\maketitle

\section{Introduction} \label{sec:intro}

\emph{Confidential computing} is an emerging model in cloud computing, which is already offered in some form by each of the three largest cloud providers~\cite{Azure-CC, GCP-CC, AWS-CC}. The primary aim of confidential computing is to protect data in use, \eg against insider threats or compromise of the underlying cloud infrastructure. Currently, the leading approach for achieving this is to use hardware-enforced Trusted Execution Environments~(TEEs).

Although there are multiple TEE technologies available, the overarching idea is the same --- to create a strong security boundary between the TEE and other software components. The misbehavior by any component outside the TEE then cannot compromise the confidentiality or integrity of the TEE. Specifically, data within the TEE can only be read or modified by code within the same TEE. TEEs often also provide \emph{remote attestation} functionality, through which a remote party can ascertain what code runs within the TEE, and use this information to make security decisions.

Current TEE technologies can be divided into two groups: those that enable the creation of one or more \emph{in-process} TEEs within an application process, and those that protect larger structures such as containers or entire VMs. An example of the former is Intel Software Guard Extensions (SGX)~\cite{anati2013innovative,Guerreiro2020} and Keystone~\cite{lee2020keystone}, which create \emph{enclaves} within an application process and protect the enclave's data from the host process, OS, and hypervisor; examples of the latter are technologies such as AMD Secure Encrypted Virtualization (SEV)~\cite{SEV}, AMD SEV Secure Nested Paging~(SNP)~\cite{SEV-SNP}, Intel Trust Domain Extensions~(TDX)~\cite{TDX}, and Arm Confidential Compute Architecture (CCA)~\cite{CCA}, which protect entire VMs against each other and the hypervisor.

However, the hardware-based nature of most modern TEEs conflicts with VM or process migration. By design, TEE memory cannot be read by the hypervisor, OS, or other software on the platform, preventing existing migration techniques to be used directly on systems containing TEEs. Whilst some newer TEE architectures support migration, this is not universally available, and there are no in-process TEE architectures that support migration.

There are several reasons why migration is important in cloud computing. From an operational perspective, the cloud provider may want to move VMs to different physical machines to reduce the number of active machines. From a security perspective, the cloud provider may need to restart specific physical machines to apply firmware security updates, \eg to defend against newly-identified side-channel attacks against the TEE~\cite{spectre,meltdown,foreshadow,cache-side-channel,plundervolt,Schaik2020SGAxeHS}. Finally, the tenants (customers) of the cloud provider may want to move their workloads to a different provider. In these scenarios, the inability to migrate a process that uses a TEE could lead to operational inefficiencies or, in the worst case, data loss if the physical machine must be restarted.

In light of this, several TEE architectures have announced native support for migrating TEEs, namely AMD SEV~\cite{SEV}, SEV-SNP~\cite{SEV-SNP} and Intel TDX~\cite{TDX}. Some prior work~\cite{Park2016,park2019emotion,Gu2017} has proposed approaches to support migration of VMs with Intel SGX enclaves.  However, these TEEs are \emph{VM-based}, not in-process.

Compared to VM migration, in-process TEEs are harder to migrate because they are not designed with migration in mind.
There have been several proposals to \emph{retrofit} migration functionality into existing in-process TEE architectures. Guerreiro et al.~\cite{Guerreiro2020} use Hardware Security Modules~(HSMs) to manage the cryptographic keys needed to securely migrate the data; Alder et al.~\cite{Alder2017} describe how to migrate the persistent state (\eg hardware-based monotonic counter values) that may be associated with a TEE. However, all of these require either changes to the hardware (which is likely to be impractical given the large deployed base), or they assume the TEE will reach a quiescent state before it is migrated, which limits the applicability of the technique.

We present \sys, a software-only design to \emph{retrofit} migration functionality into existing TEE architectures, whilst maintaining their expected security guarantees. The core idea is to add a minimal set of extra functionality to the TEE and then \emph{enlighten} the migration tool to make use of this functionality. Our design makes the following contributions:

\vspace{3pt}

\noindent
(1)~It enables migration of \emph{existing} in-process TEE architectures, without requiring modifications to the hardware or placing constraints on the software running within the TEE. This is challenging because it requires a software-only solution that can operate within the constraints of existing TEE architectures (\eg in Intel SGX, some critical data structures are inaccessible even from software within the enclave). Furthermore, these architectures may not have been designed with migration in mind.

\vspace{3pt}

\noindent
(2)~It allows TEEs to be interrupted, migrated, and resumed at arbitrary points in their execution, thus matching the paradigm of existing process migration tools, such as CRIU~\cite{criu}, which expect to migrate a process at any point in time.  This allows us to integrate \sys with these existing tools with minimal changes.  Our method for \emph{enlightening} these tools is itself extensible, and could be used to enable new types of process migration behavior.

\vspace{3pt}

\noindent
(3)~It involves the TEE in the migration operation, resulting in a type of \emph{cooperative migration}. This gives TEE application developers the ability to specify flexible stateful policies to govern migration. Examples of such policies may be to limit the number of times a particular TEE is migrated, or to migrate only a subset of the TEE's memory.

\vspace{3pt}

\noindent
As a proof of concept, and for our performance evaluation, we have implemented \sys for Intel SGX. Through micro and macro benchmarks, we show that migration latency increases linearly with the size of the TEE, and that the overhead is dominated by TEE system operations, mainly creation and termination of TEEs.

\section{Migration is the Cloud} \label{sec:motivation}

Liveness for a large number of cloud services replies on the availability of the machines on which the service is deployed~\cite{Azure-availability,GCP-availability,AWS-availability}.
This architecture runs counter to the deployment philosophy of many cloud providers.
Cloud providers consider a single machine to be expendable and instead define large availability zones where machines within a single availability zone may become simultaneously unavailable but machine in different availability zones will not be unavailable at the same time~\cite{Azure-availability-zone,GCP-availability-zone,AWS-availability-zone}.
In these situation migrating the application from a machine which the cloud provider will shutdown in the near future allows the services to maintain a high level of availability.
Therefore, it is important that TEEs also support migration, especially as confidential computing are gaining popularity amongst cloud providers.
In this section, we first describe \emph{VM-based} TEEs that support migration and their drawbacks, and then discuss the state-of-the-art solution for \emph{in-process} TEEs.

\mypar{TEE VM migration}
AMD has enabled its TEE extensions SEV and SEV-SNP with hardware support for live migration~\cite{SEV,SEV-SNP}.
The hardware extensions provided by AMD SEV are called AMD Secure Processor (SP) which manage the encryption keys required for live migration.
During TEE VM migration, the destination SP attests itself to the source SP and once the attestation is verified, the source SP sends the key securely to the destination SP.
After the verifications is completed successfully the source SP sends the encrypted TEE VM pages to the destination SP, which uses the key sent by the source SP to decrypt the pages and copy them into the destination TEE VM.

AMD SEV-SNP improves upon the AMD SEV model by introducing \emph{migration agents}.
Migration agents oversees the enforcement of migration policies removing the requirement for the VM to maintain its own migration policy.

Intel announced live migration support for their unreleased TEE extension, TDX~\cite{TDX}.
TDX creates TEE VMs, called \emph{trust domains} (TDs), and similar to SEV-SNP's migration agent, TDX utilizes an entity called \emph{migration TD} when migrating a TD.
The source and migration TDs conduct a mutual remote attestation and negotiate a key to encrypt the contents of the migrating TD.

Although Intel's current TEE architecture, SGX, does not natively support migration, several attempts had been made to migrate SGX-enabled VMs in the literature.
The first attempt was by Park \etal where they proposed a design that supports live migration of SGX-enabled VMs~\cite{Park2016}.
They pointed out several problems of migrating such VMs, including the secure migration of enclave memory.
The proposed idea was to introduce a new set of CPU instructions to enable live migration.
In their follow-up work~\cite{park2019emotion}, the authors implemented the proposed instruction using OpenSGX~\cite{jain2016opensgx}, a fully functional Intel SGX emulator based on QEMU.

In parallel work, Gu \etal~\cite{Gu2017} utilize a control thread to securely copy an enclave's state from within the TEE.
The system uses a scheme called ``two-phase checkpointing'', which requires an enclave to reach a checkpoint before migrating, to ensure data consistency of migrated enclaves.
The authors implemented their system a variant of the KVM that provides support for Intel SGX in applications, guest OSs, and the KVM itself.

Unfortunately, these attempts cannot migrate a process running within an operating system without the coordination of application being aware and adding logic which coordinates with the migration coordinator.
This results with TEEs losing a considerable amount of utility when run on a machine within a cloud data center.
Additionally, the aforementioned systems cannot migrate a process running within the TEE without migrating the entire VM.
This increases the latency and also adds strain to the cloud provider's network bandwidth, as a larger amount of data must be migrated.
This motivates migrating \emph{in-process} TEEs without the need of migrating the entire underlying VM.

\mypar{In-process TEE migration}
To the best of our knowledge, TEEnder~\cite{Guerreiro2020} is the only work that aims to enable migration for in-process TEEs, specifically Intel SGX.
TEEnder uses Hardware Security Modules (HSMs) to encrypt and decrypt SGX enclave data during migration.
The main motivation behind using HSMs is the recent findings of security vulnerabilities surrounding SGX remote attestation~\cite{lee2017hacking,schwarz2019zombieload,Swami2017IntelSR,Schaik2020SGAxeHS}.
TEEnder realizes this by integrating enclave applications with HSMs and implementing an infrastructure that utilizes HSMs to provide enclaves with migration capabilities.
However, the usage of HSMs, although motivated clearly, will decrease the deployability of the system as well as the performance of migration.

Reflecting the challenges shown above, a design for migrating processes containing TEEs must fulfil the following requirements:

\begin{enumerate}
    \item[\textbf{R1}] The migration functionality must be \emph{retrofitted} into existing TEE architectures.
    \item[\textbf{R2}] The design must maintain the existing security guarantees provided by the TEEs.
    \item[\textbf{R3}] The TEE developer must be able to define stateful policies that govern the migration.
\end{enumerate}

We present an overview of our design in the next section, and show how it meets these requirements in Section~\ref{sec:design}.

\section{System Overview}
\label{sec:system_overview}

\begin{figure*}[tb]
    \centering
    \includegraphics[width=.95\textwidth]{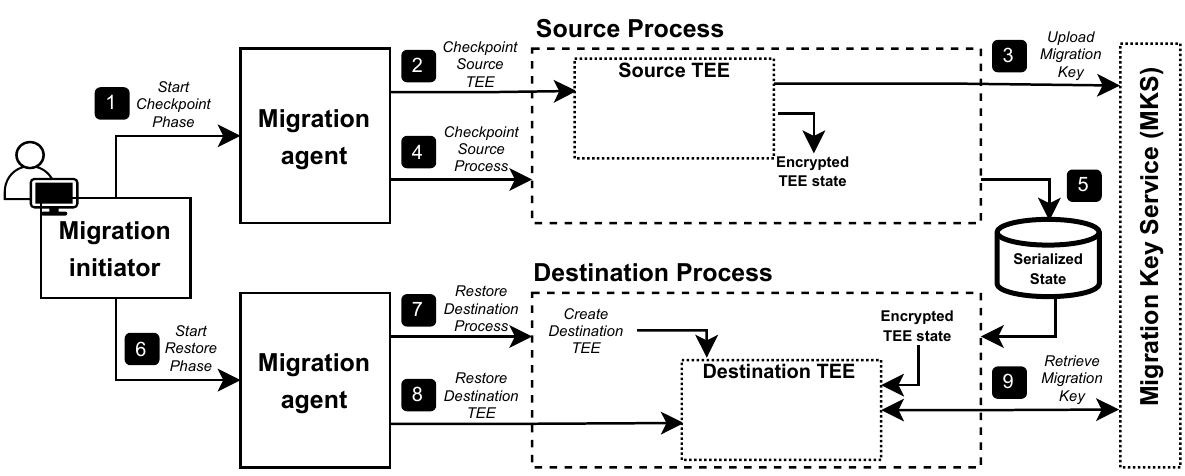}
    \caption{Overview of the main components \sys and the interactions between them.}
    \label{fig:Overview}
  \end{figure*}

Figure~\ref{fig:Overview} shows an overview of the main entities involved in \sys, and the interactions between them.

The migration begins when the \emph{migration initiator} decides that a process should be migrated from the node on which it is currently running (the \emph{source}) to another node (the \emph{destination}).
The source process contains at least one TEE that must be included in the migration.
The migration can be broadly divided into two phases: \emph{\checkpoint} (\S\ref{sec:system:checkpoint}) and \emph{\restore} (\S\ref{sec:system:restore}).

\subsection{\Checkpoint phase}
\label{sec:system:checkpoint}

The first step in the \checkpoint phase is to pause the execution of the source process by stopping all currently executing threads, including those within the TEE.
With the execution paused, the state of the source process can be serialized and saved.
There are several existing tools, such as CRIU~\cite{criu}, that can be used to serialize and save the state of the source process. 
We refer to this type of tool as the \emph{migration agent} on the source (and destination) node.

However, since the migration agent is running outside the TEE, it cannot directly read the TEE's memory, as would be required to perform the migration.
Using an unmodified migration agent on a process containing a TEE typically results in an access violation.
\sys therefore requires a small amount of additional functionality to the TEE, through the inclusion of a software library.
This functionality can be used to request the TEE to serialize its own state, encrypt it, and write the output to the memory of the source process.
The cryptographic key used to encrypt (and subsequently decrypt) the TEE's state cannot be revealed to the source process, so it is securely transferred to a trusted \emph{migration key service}.

The existing migration agent can be \emph{enlightened} to i) recognize that the source process contains a TEE, and ii) to use the additional functionality added to the TEE to perform the migration.
The output of the \checkpoint phase is the serialized state of the source process, including the encrypted state of one or more TEEs.
This state is then transferred to the destination host.

\subsection{\Restore phase}
\label{sec:system:restore}

The \restore phase begins when the \destination node receives the state from the \checkpoint phase.
This state is used to recreate the saved process (now referred to as the \emph{destination process}), using a migration agent on the destination node (\eg CRIU).
Using information from the saved process state, the migration agent also creates one or more fresh TEEs in the destination process, corresponding to the TEEs that were paused in the source process.

However, the migration agent does not have the decryption keys for the encrypted TEE state and cannot write directly to the memory of the destination TEE.
Similar to the \checkpoint phase, \sys delegates the task of restoring the TEEs' internal state to the TEEs themselves.
Specifically, the additional functionality added to the TEE can also be used to restore a previously-saved TEE state onto a newly-initialized TEE.

The destination TEE securely retrieves the decryption keys from the migration node, using remote attestation to demonstrate that it is the correct type of TEE running the expected code.
Once the keys have been retrieved, the destination TEE decrypts the saved state and transforms itself by overwriting its own heap, stack, and other data structures with those from the restored state.
Once the migration has been completed, this restored TEE can continue operating from exactly the same point at which the original source TEE was paused.

\section{Design Challenges \& Solutions}
\label{sec:design}

This section discusses the main challenges arising from the requirements in Section~\ref{sec:motivation}, and approaches used to overcome these in \sys.

\subsection{Retrofitting migration to existing TEEs}
\label{sec:design:retrofit}

The requirement to support existing TEE architectures, which may already be widely deployed, precludes the possibility of modifying TEE hardware (\eg as done by Park et al.~\cite{Park2016}) and necessitates a software-only approach.
Since only the software running within the TEE can read and write TEE memory, the \checkpoint and \restore functionality must be provided from within the TEE.
As explained in Section~\ref{sec:system_overview}, \sys adds these functionalities by including an additional software library within the TEE.
In future, this library could be included by default in the TEE development frameworks (\eg Open Enclave~\cite{oe}).
Performing the migration from within the TEE raises specific challenges in both the \checkpoint and \restore phases.

\mypar{Arbitrarily pausing TEEs} Halting a TEE without modifying TEE hardware is a challenge. 
One way of doing this is to wait until the TEE reaches a certain point in its execution and exit (\eg as done by Gu \etal~\cite{Gu2017}).
However, this requires the TEE to be aware of the migration and thus may not fit requirements of certain migration agents (\eg CRIU requires processes to be arbitrarily paused).
This means that \sys must be able to halt a TEE at any point in time.
\sys realizes this by using \emph{interrupts}, which is widely supported across different TEE architectures.

\mypar{Operating within the TEE} The \checkpoint functionality is invoked once the source TEE has been paused, and therefore must carry out its operations without affecting the state of the source TEE.
In addition, the working state of this operation must not be included in the saved state of the TEE.
The exact techniques used to overcome these challenges will vary by TEE technology (\eg we describe our implementation for Intel SGX in Section~\ref{sec:impl}), but in general will require the \checkpoint operation to use its own stack and heap memory.
The \restore functionality faces similar challenges that it must decrypt and process the saved state within the fresh destination TEE and then overwrite the state of that TEE (stack and heap) with the restored state.
This likewise means that the \restore operation must use its own reserved memory within the TEE.

\mypar{Understanding TEE memory} The next challenge is that the \checkpoint and \restore operations within the TEE must understand the TEE's memory map.
For example, the TEE's memory might contain control structures that cannot be read or written even by software running within the TEE (\eg the Thread Control Structures in Intel SGX).
The migration operations must take care to avoid these memory regions.
Additionally, \sys may need to use architecture-specific techniques to infer the values held in these control structures in the source TEE and to correctly set these values in the destination TEE.

\subsection{Maintaining TEE security}
\label{sec:design:tee_security}

When migrating TEEs, \sys needs to ensure that existing TEE security guarantees still hold.
Specifically, the only change to the security model is that the TEE can be migrated.
As explained in Section~\ref{sec:system_overview}, the \checkpoint operation encrypts the TEE's memory before writing it outside the TEE.
Specifically, an authenticated encryption scheme, such as AES-GCM, must be used so that the integrity of the saved state can be verified by the destination TEE. 
We refer to the symmetric encryption/decryption key as the \emph{migration key}.

\mypar{Secure key transfer} The migration key cannot be directly exported with the saved state --- it must instead be securely transferred to the destination TEE.
If the source and destination TEEs were running concurrently, they could use existing remote attestation functionality to mutually attest each other, establish a secure channel, and securely transfer the migration key.
However, requiring both TEEs to be running concurrently would severely limit the applicability of the approach.
For example, existing process migration tools such as CRIU~\cite{criu} operate strictly sequentially: the \checkpoint phase is completed before the \restore phase begins.
Requiring concurrently running source and destination TEEs would also preclude the possibility of \emph{self-migration}, where the source and destination are the same physical node.
This would be used to allow the node to be restarted \eg to install security firmware updates.

To overcome this challenge, \sys makes use of a new \emph{migration key service} (MKS), which serves as a trusted intermediary and key escrow service between the source and destination TEEs.
Specifically, once the source TEE has generated the migration key, it establishes a secure channel with the MKS and sends the migration key.
The \restore operation in the destination TEE retrieves the migration key from the MKS and uses it to decrypt the TEE state.

\begin{figure}[t]
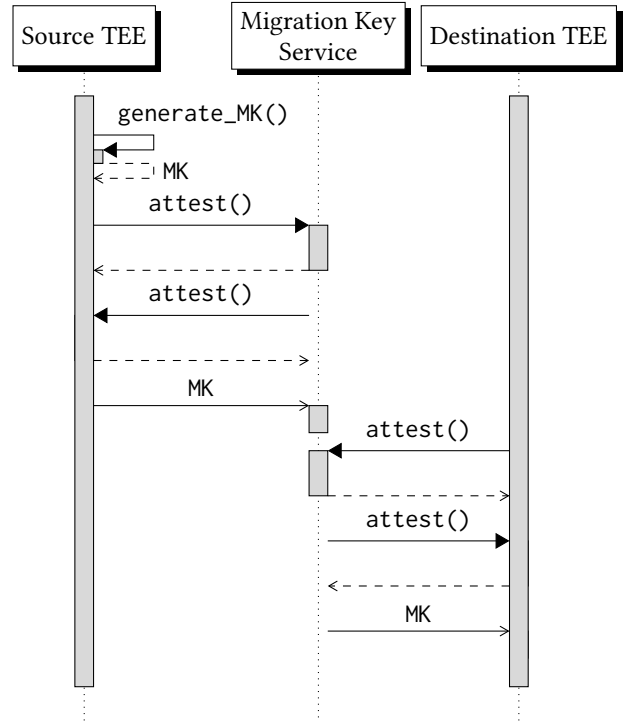

\newcommand{\generatemk}{\texttt{generate\_MK()}}
\newcommand{\attest}{\texttt{attest()}}
    \centering
    \begin{sequencediagram}
        \newthread{s}{\shortstack{Source TEE}}
        \newinst[1]{mks}{\shortstack{Migration Key\\Service}}
        \newthread{d}{\shortstack{Destination TEE}}
        \begin{callself}{s}{\generatemk}{\mk}
        \end{callself}
        \begin{call}{s}{\attest}{mks}{}
        \end{call}
        \begin{call}{mks}{\attest}{s}{}
        \end{call}
        \begin{messcall}{s}{\texttt{MK}}{mks}
        \end{messcall}
        \prelevel
        \begin{call}{d}{\attest}{mks}{}
        \end{call}
        \begin{call}{mks}{\attest}{d}{}
        \end{call}
        \begin{messcall}{mks}{\texttt{MK}}{d}
        \end{messcall}
    \end{sequencediagram}

    \caption{Secure transfer of the migration key (\mk) from source to destination TEE via the Migration Key Service. All communication takes place via secure channels.}
    \label{fig:migration_tee}
\end{figure}

The MKS is a relatively simple store-and-forward helper service, for which there are various possible implementations.
One possibility is to implement the MKS using another TEE, either on the source, destination, or another node.
As shown in \cref{fig:migration_tee}, the source TEE attests the MKS to ensure that the migration key is only transferred to a trustworthy MKS, and the MKS attests the source TEE to verify the provenance of this key.
Subsequently, the MKS attests the destination TEE to ascertain that it is the correct type of TEE (\eg the same as the source TEE), and the destination TEE attests the MKS to again verify the provenance of this key.
This provides the same security guarantee as the direct key transfer between source and destination TEE.

\mypar{Fork and roll-back attacks} In addition to secure key transfer, the design must also ensure that the migration functionality itself cannot be used to mount attacks such as a fork or roll-back attack~\cite{Alder2017}.
Specifically, we need a mechanism to ensure that the source TEE cannot continue running after the \checkpoint operation has been completed, as this could lead to a fork attack with multiple copies of the same TEE running.
To prevent this, \sys blocks the source TEE from being resumed before the output from the \checkpoint operation is released.
The precise implementation is architecture-dependent, but can always be achieved through minor modifications of the TEE's software.
We also require a mechanism to prevent one saved state being restored to multiple destination TEEs (another type of fork attack) or being restored more than once (a roll-back attack).
This cannot be prevented by changes within the TEE, so \sys requires the MKS to only release the migration key to a single destination TEE.

\subsection{Supporting policy-governed migration}
\label{sec:design:policy}

The design must provide mechanisms that TEE developers can use to define and enforce policies to govern the migration of the TEE.
\sys does not define any specific policies, but aims to provide as much flexibility as possible to TEE developers.
Specifically, \sys enforces policies by calling a developer-defined function (which call additional functions) during the \restore phase before allowing the TEE to resume operation.
Since the TEE state has already been decrypted and put into place, this function can be stateful and can inspect the full state of the TEE.
The return value of this function indicates whether the TEE should be allowed to resume operation.
This ensures that every restore operation is visible to the TEE.
We sketch two example policies to illustrate the use of this mechanism.

\mypar{Limited number of migrations} One example policy could be to limit the number of times a specific TEE can be migrated.
This may be useful in cases where the TEE developer would like to allow the cloud provider to migrate the TEE a limited number of time, but may be concerned that an excessive number of migrations might leak information from the TEE (\eg through side-channel attacks).
To implement this, the developer would define a counter variable in the TEE's memory and decrement this on each successful \restore operation.
When the counter reaches zero, the function would indicate that the TEE should not be permitted to resume. 

\mypar{Clearing caches upon migration} As another example, the policy might not need to govern whether the TEE can be resumed, but rather define a set of actions that must be performed after each migration.
For example, the TEE might contain node-specific state (\eg some type of cache of local sessions) that should be invalidated if the TEE is migrated.
Again this can be achieved by calling a developer-defined function that clears/resets this part of the state whenever the TEE is restored.

\section{Implementation} \label{sec:impl}
This section describes \syssgx, an implementation of \sys for Intel Software Guard Extensions.
Our implementation is based on the Open Enclave SDK~\cite{oe} v0.16.1, but could be applied to any other SGX SDK.
We first describe the specific steps required for migrating an Intel SGX enclave (\S\ref{sec:enclave_migration}) and then discuss how we integrated these into the \criu~\cite{criu} process migration tool (\S\ref{sec:sys_impl}).

\subsection{Enclave Migration} \label{sec:enclave_migration}

In practice, migrating an SGX enclave requires some additional preparation before the \checkpoint phase and some additional cleanup after the \restore phase.
We therefore describe this process in terms of the following four phases:
\begin{enumerate}
    \item \textbf{Preparation Phase}: Initializes variables used for migration.
    \item \textbf{\Checkpoint Phase}: Collects, encrypts, and exports all necessary data from the source enclave.
    \item \textbf{\Restore Phase}: Imports, decrypts, and restores all data to the destination enclave.
    \item \textbf{Cleanup Phase}: Cleans up data structures (e.g., buffers) used for migration.
\end{enumerate}

Below, we describe each phases in detail.

\begin{figure*}[t]
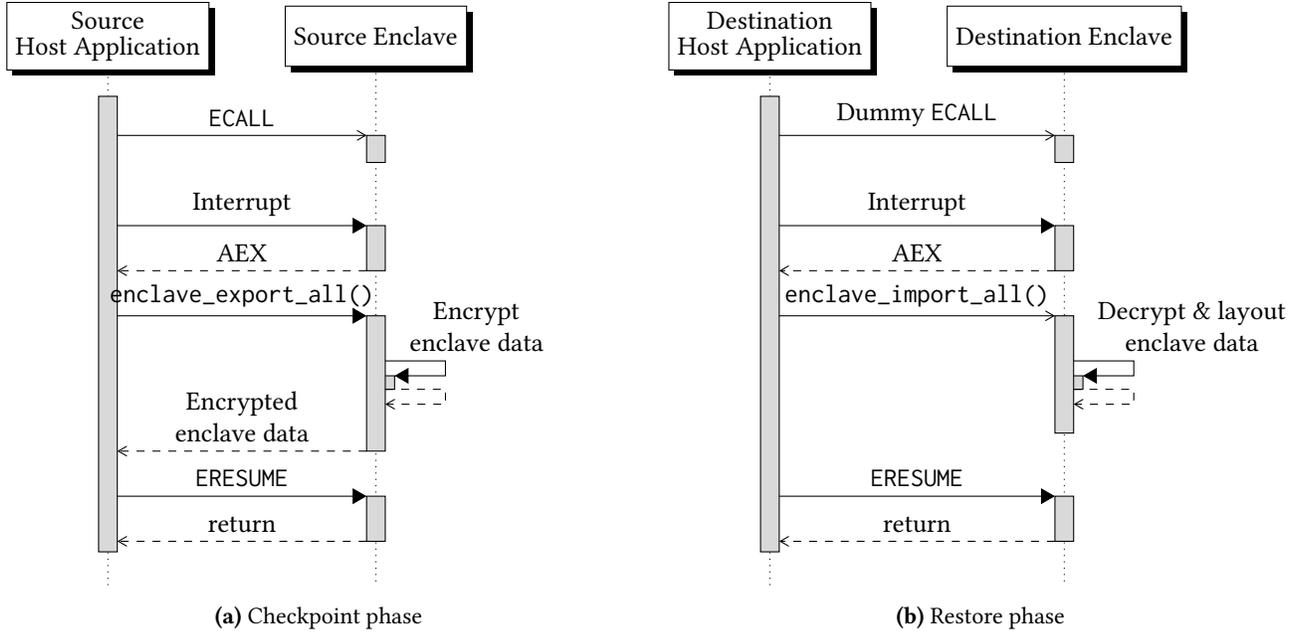

    \begin{subfigure}{0.49\textwidth}
    \begin{sequencediagram}
        \newthread{sh}{\shortstack{Source\\Host Application}}
        \newinst[1]{se}{Source Enclave}

        \begin{messcall}{sh}{\ecall}{se}
        \end{messcall}
        \begin{call}{sh}{Interrupt}{se}{AEX}
        \end{call}
        \begin{call}{sh}{\texttt{enclave\_export\_all()}}{se}{\shortstack{Encrypted\\enclave data}}
        \begin{callself}{se}{\shortstack{Encrypt\\enclave data}}{}
        \end{callself}
        \end{call}
        \begin{call}{sh}{\texttt{ERESUME}}{se}{return}
        \end{call}
    \end{sequencediagram}
    \caption{\Checkpoint phase}
    \label{fig:em_dump_flow}
    \end{subfigure}
    \begin{subfigure}{0.49\textwidth}
    \begin{sequencediagram}
        \newthread{dh}{\shortstack{Destination\\Host Application}}
        \newinst[1]{de}{Destination Enclave}

        \begin{messcall}{dh}{Dummy \ecall}{de}
        \end{messcall}
        \begin{call}{dh}{Interrupt}{de}{AEX}
        \end{call}
        \begin{messcall}{dh}{\texttt{enclave\_import\_all()}}{de}
        \begin{callself}{de}{\shortstack{Decrypt \& layout\\enclave data}}{}
        \end{callself}
        \end{messcall}
        \begin{call}{dh}{\texttt{ERESUME}}{de}{return}
        \end{call}
    \end{sequencediagram}
    \caption{\Restore phase}
    \label{fig:em_restore_flow}
    \end{subfigure}
    \caption{Enclave migration flow}
    \label{fig:em_flow}
\end{figure*}

\subsubsection{Preparation Phase} \label{sec:em_prep}

The preparation phase is run asynchronously, any time before the enclave is to be migrated.
This can either be integrated into the enclave's initialization sequence, or called via a separate \ecall.
In this phase, the enclave generates a migration key and initializes the encryption context (\ie the data structures used by the cryptographic library).
It also retrieves the memory addresses and sizes of the enclave's stack, heap, and data sections, as well as the address and size of the SGX-specific State Save Area (SSA).
These values are stored within the enclave in preparation for migration.

\subsubsection{\Checkpoint Phase} \label{sec:em_dump}

To initiate the \checkpoint phase, the host should interrupt all enclave threads and call a newly-added \ecall (called \texttt{enclave\_export\_all}).
In this \ecall, the host provides a pointer to a memory buffer outside the enclave, into which the encrypted state should be written.
The enclave then performs several steps in order to serialize, encrypt, and save its own state.

\mypar{Halting enclave threads}
\syssgx uses interrupts to pause TEEs.
In Intel SGX, interrupting a thread that is within the enclave causes an event called an Asynchronous Exit (AEX).
When an AEX occurs, the CPU saves the thread's state in the enclave's State Save Area (SSA).
This saved state includes the Thread Local Storage (TLS), current instruction pointer, stack and base pointers, and other register values (within a data structure called GPRSGX).
The CPU then clears the registers and jumps to a predefined code location outside the enclave. 

The host process (\ie the source process) can send a signal to each thread within the enclave to cause an AEX.
However, the default behavior of most SGX frameworks is for the thread to immediately re-enter and resume executing within the enclave after the interrupt has been handled.
The host process therefore needs to take additional steps to prevent the thread re-entering the enclave.
A simple solution would be to cause the thread to sleep or spin in an infinite loop.
However, all enclave threads jump to the same code address outside the enclave when they are interrupted.
At least one thread will be required to perform the \checkpoint operations within the enclave, and this thread might also be periodically interrupted.
If this thread is interrupted, it should not be held outside the enclave.

The host process can overcome the above issues using three separate flags: \texttt{IS\_HOST\_MIGRATING}, \texttt{HAS\_ENTERED\_ENCLAVE}, and \texttt{ALLOW\_ERESUME}.
All three flags start in the unset state.
The host process first sets the \texttt{IS\_HOST\_MIGRATING} flag to indicate that it is going to halt the enclave thread.
This flag causes the interrupted enclave threads to be kept in an infinite loop, which is conditioned to break when the \texttt{ALLOW\_ERESUME} flag is set.
Finally, the \texttt{HAS\_ENTERED\_ENCLAVE} flag is set immediately before the migration thread enters the enclave.
This flag prevents any new threads from being captured in the infinite loop, whilst still holding the previously-captured threads.

Once the migration thread has entered the enclave (via the \texttt{enclave\_export\_all} \ecall), it could set an in-enclave flag to prevent any other \ecall{s} being made.
It could also prevent any threads being resumed while the migration is in progress by saving and overwriting the saved instruction pointer values in the SSA.
If any thread did resume operation after this point, it would cause the enclave to crash.

\mypar{Exporting enclave data}
With the enclave paused, the next step is to encrypt and export the enclave's data.
The data to be are exported are: (1) data section, (2) heap section, (3) stack, and (4) SSA.
The enclave data are serialized by concatenating all the above in the order shown.
We do not export code data because we assume both the source and destination enclaves are running the same code.

Additionally, every SGX enclave includes one or more Thread Control Structures (TCS), which cannot be read or written even by code running within the enclave.
Each TCS contains a Current State Save Area (CSSA) value, which indicates how many threads have been interrupted and are now outside of the enclave.
Since we are using interrupts to halt enclave threads, we need to infer the CSSA value.
We could employ the same method proposed by Gu \etal~\cite{Gu2017}, which introduces a software monitor that keeps track of how many threads have entered and exited the enclave.

Recall that enclave data must be encrypted before it can be written outside the enclave (Section~\ref{sec:design:tee_security}).
We use AES in GCM mode (\ie authenticated encryption) with a 256~bit key for this purpose.
We implemented this using both the cryptographic libraries supported by Open Enclave: mbedTLS and OpenSSL.
Once encrypted, that data is written to the specified buffer outside the enclave.

\subsubsection{\Restore Phase} \label{sec:em_restore}

The \restore phase also consists of several steps. 
First, the destination enclave must be created and initialized.
Second, \syssgx must create one or more placeholder threads within the destination enclave, corresponding to the threads that were interrupted from the source enclave.
Third, the host process initiates the \restore process by calling a new \ecall (\texttt{enclave\_import\_all}).
This \ecall takes the pointer to the encrypted enclave data received from the source process.
Finally, the restored enclave threads resume execution from the point at which they were interrupted.

\mypar{Creating placeholder threads}
Placeholder threads are essentially enclave threads with an infinite loop.
There are several reasons to why we need to create placeholder threads in the destination enclave.
Firstly, we need a pre-allocated memory region so that we can lay out the exported enclave data back to its original location.
Secondly, this causes the CPU to increase the CSSA value in the TCS.
This technique overcomes the limitation of not being able to write to the TCS from software, and can thus be used to set the correct CSSA values.
\syssgx therefore creates a placeholder thread in the destination enclave for each thread that was interrupted from the source enclave.

\mypar{Restoring enclave data}
The main challenge during the restoration phase is to ensure that the enclave data is restored to the correct location.
For security reasons, the saved state must be decrypted within the destination enclave.
The data structure used by the cryptographic library (\ie the decryption context) is typically allocated on the heap.
This creates a problem when we restore the heap section, because we may potentially overwrite the decryption context when we are decrypting and restoring the subsequent section.
To overcome this issue, \syssgx places the decryption context in the global data section and intentionally avoids that area during the restoration process.
This is possible because the code is identical between the source and destination enclaves, allowing the encryption and decryption contexts to be allocated in the same memory location.

\subsubsection{Cleanup Phase} \label{sec:em_cleanup}
Both the source and destination processes carry out a cleanup phase after they have completed their respective roles in the migration.
On the source node, after the migration \ecall has returned, the host process can tear down the enclave.
Alternatively, if allowed by the enclave developer, the host process may be allowed to resume the enclave by setting the \texttt{ALLOW\_ERESUME} flag.
This is essentially ``forking'' the enclave, which may be desirable in some circumstances.
On the destination node, after the migration process has been completed and the enclave resumed, the buffer that was used to store the encrypted enclave data can be freed.

\subsection{Integrating \syssgx into CRIU} \label{sec:sys_impl}
In this section, we describe how enclave migration can be integrated into existing process migration tools.
Specifically, we integrate \syssgx into \criu~\cite{criu}.
The following subsections describe this integration in each of the four phases discussed above.
These descriptions are written with respect to a source process that contains a single enclave, but would equally apply if the source process were running multiple enclaves. 

\subsubsection{Preparation Phase} \label{sec:sys_prep}
In addition to the steps described in Section~\ref{sec:em_prep}, the source process must perform several steps.
First, it must close file descriptors to \texttt{stdin}, \texttt{stdout}, \texttt{stderr}, and \texttt{/dev/sgx}, as they cannot be migrated by \criu.
Second, it must allocate a memory buffer to store the encrypted enclave memory, the address of which is passed as a parameter when calling \texttt{enclave\_export\_all}.
Finally, it creates a file containing code pointers to several functions within the source process, which will be used in both the \checkpoint and restoration phases.

\subsubsection{\Checkpoint Phase} \label{sec:sys_dump}

This phase begins when the migration initiator instructs the migration agent (in this case, \criu) to migrate a specific process (in this case, the source process). 
\criu process transitions the source process into a ``seized'' state and begins to determine what must be migrated.
In order to do this, \criu injects a piece of code (called the \emph{parasite code}) into the source process.
The parasite code allows \criu to gain access to resources held by the source process, such as file descriptors and threads.
However, by design, even this parasite code cannot read the memory of the SGX enclave.

\mypar{Extending \criu}
\syssgx therefore extends the parasite code to work with the source enclave in order to achieve the migration.
Specifically, we extend the parasite code to call the \texttt{enclave\_export\_all} \ecall.
Although it may be possible to make this \ecall from outside the source process (\eg for the migration agent to make this \ecall), this would have to deal with several address translations.
A more natural approach is for the injected parasite code to make this \ecall, since the parasite code i) has direct access to variables in the source process, such as the enclave context, and ii) can directly invoke functions from the source process's address space, such as the \ecall.

\mypar{Determining function addresses}
However, the addresses of the \texttt{enclave\_export\_all} \ecall and other enclave functions are not deterministic, due to memory address randomization.
To overcome this issue, our enlightened version of \criu looks up the necessary function pointers in the file created in the preparation phase (\S\ref{sec:sys_prep}).
Specifically, the parasite code opens the file from a predefined location, retrieves the function pointers, and invokes each of the functions sequentially.

\mypar{Passing parameters}
Since this approach does not allow \criu to pass arguments to any of the functions, the function pointers in the file should point to wrapper functions that do not take any parameters.
These wrapper functions could in turn call other functions for which the parameters have been determined in advance.
A specific example is the \texttt{enclave\_export\_all} \ecall, which must include the address of the memory buffer outside the enclave into which the encrypted state will be written.
This parameter can be determined when the buffer is allocated (\ie in the preparation phase).
\criu calls a wrapper function with no parameters, which in turn calls this \ecall with the pre-prepared parameter value.

\mypar{Further extensibility}
This approach is extensible in that the application developer can provide additional functions for \criu to invoke before starting the migration.
Apart from enabling cooperative migration of TEEs, this could also be used to support other types of extended functionality \eg if the source process needs to perform some non-standard clean-up operations or close some resources before being migrated.

\mypar{Overall \checkpoint process}
The interaction between the source process and the source enclave is shown in Figure~\ref{fig:em_dump_flow}.
Putting it all together, the \checkpoint process proceeds as follows:
When \criu seizes the source process, this will interrupt and pause all of the source process's threads.
Specifically, this will cause an AEX for any threads within the enclave, as required.
Once the process is seized, \criu injects the parasite code into the source process and invokes it.
The parasite code then calls the \texttt{enclave\_export\_all} \ecall, which as explained above, causes the enclave to serialize, encrypt, and write its state to the memory of the source process.
This memory will then be included in the saved process image output by \criu, alongside with the rest of the source process's memory.
Finally, after all the data has been exported, the parasite code destroys the enclave.
The migration agent can then transfer the saved process image to the destination node.

\subsubsection{\Restore Phase} \label{sec:sys_restore}
The \restore phase begins once the migration initiator has transferred the process image to the destination node and invoked the migration agent (\criu) on the destination node.
The \criu process on the destination node uses this process image and essentially \emph{transforms} itself into the restored destination process.
In order to restore the process memory, \criu uses another separate piece of code called the \emph{restorer blob}.
As with the parasite code in the \checkpoint phase, \syssgx extends the \criu restorer blob to handle restoration of the enclave.
The interaction between the destination process and the destination enclave is shown in Figure~\ref{fig:em_restore_flow}.

\mypar{Restoring a process with \criu}
\criu first restores resources that do not require the restorer blob, such as threads.
This includes the enclave threads that were exported during the \checkpoint phase.
\criu then injects the restorer blob into its own process memory and uses this blob to restore the memory from the saved process image.

\mypar{Restoring an enclave with \criu}
Once the destination process's memory has been restored, the restorer blob needs to create and restore the destination enclave.
\syssgx uses a similar mechanism as for the parasite code in the \checkpoint phase in that the restorer blob accesses a file at a predefined location and reads a list of function pointers to invoke.
As explained above, this file of function pointers was created during the preparation phase.
Since the destination process does not yet contain an enclave, the restorer blob first invokes the function to create an enclave and the \ecall to initialize important variables (e.g., encryption context) within the enclave.
It then creates one or more placeholder enclave threads, corresponding to the number of enclave threads that were interrupted.
Each of these threads is then interrupted and destroyed once outside the enclave.
The restorer blob then invokes the \texttt{enclave\_import\_all} \ecall (via the corresponding wrapper function).
As explained in Section~\ref{sec:em_restore}, this \ecall restores the exported enclave data to the correct location.

\subsubsection{Cleanup Phase} \label{sec:sys_cleanup}
In addition to the steps described in Section~\ref{sec:em_cleanup}, the injected restorer blob must be removed from the destination process's memory.
Finally, the restored process is released from its seized state and starts resuming its execution.
Once the restored process is resumed, the restored enclave thread can re-enter the enclave and continue execution.

\section{Evaluation} \label{sec:eval}
We evaluated \syssgx to understand the overheads and performance impact of migrating in-process TEEs.
In \S\ref{sec:exp-setup} we first describe the settings of the environment and benchmarks used for the evaluation.
Next, we report the throughput of an application running in an enclave and the performance impact of when it is migrated (\S\ref{sec:throughput_eval}).
Then, we show the overall latency of migrating both the host application and enclave (\S\ref{sec:overall_latency}).
Next, we report the latency of migrating just the enclave (\S\ref{sec:enclave_migration_latency}) and the host application (\S\ref{sec:host_migration_latency}).
Finally, we consider the cryptographic operations used in \syssgx (\S\ref{sec:latency_vs_crypto_lib}).

\subsection{Experimental Setup}
\label{sec:exp-setup}
We conducted our experiments using a Microsoft Azure Confidential Computing DC4ds v3 series VM.
The VM had 4 Intel Xeon Platinum 8370C vCPUs (Icelake, 2.80 GHz) with 32 GB of assigned memory of which 16 GB could be used by the SGX TEE's enclave page cache (EPC).
We ran Ubuntu 18.04 and installed the Intel SGX DCAP Driver version 1.33.2.
To remove disk I/O latency from the critical path, we used a RAM disk to store images created by \syssgx's \checkpoint phase.

\mypar{Implementation}
We built our implementation of \syssgx in $913$ lines of C/C++ code of which $96$ lines were additions or modifications to the Open Enclave SDK and $204$ lines were additions or modifications to \criu.
\syssgx was integrated into \criu version 3.15 and Open Enclave version 0.16.1.
All cryptography was performed using either mbedTLS version 2.16.10 or OpenSSL version 1.1.1k and \syssgx used OpenSSL unless otherwise specified.

\mypar{Benchmarks}
The throughput evaluation used the \emph{SmallBank} benchmark~\cite{Alomari2008} which models a bank with $1,000$ accounts.
The clients perform one of five randomly selected transactions that either: deposit funds; transfer funds; withdraw funds; check an account's balance; or amalgamate two accounts.
We implemented the SmallBank benchmark with SQLite version 3.34.1.
We ran the benchmark for 10 seconds allowing it to reach a steady state before migrating it to a different process, and then ran it for another 10 seconds ensuring the migrated process also reached a steady state.

\subsection{Throughput} \label{sec:throughput_eval}

\begin{figure}[t]
    \centering
    \begin{subfigure}{0.47\textwidth}
    \centering
\ifdefined \showFullFigure
    \includegraphics[width=\columnwidth]{figs/sqlite_throughput_full.pdf}
\else
    \includegraphics[width=\columnwidth]{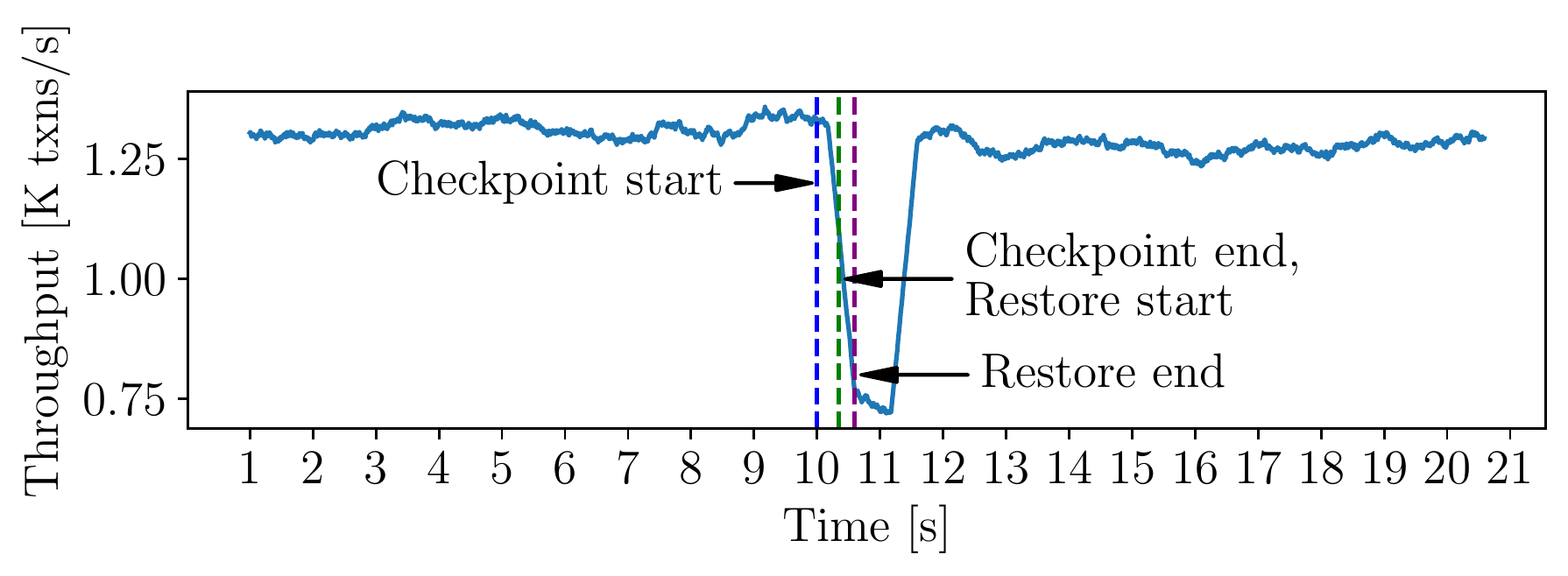}
\fi
    \caption{The blue line is the average number of transactions in the previous second. The vertical lines show when certain migration operations occurred.}
    \label{fig:throughput_eval}
    \end{subfigure}

    \begin{subfigure}{0.47\textwidth}
    \centering
\ifdefined \showFullFigure
    \includegraphics[width=\columnwidth]{figs/sqlite_throughput_hist_full.pdf}
\else
    \includegraphics[width=\columnwidth]{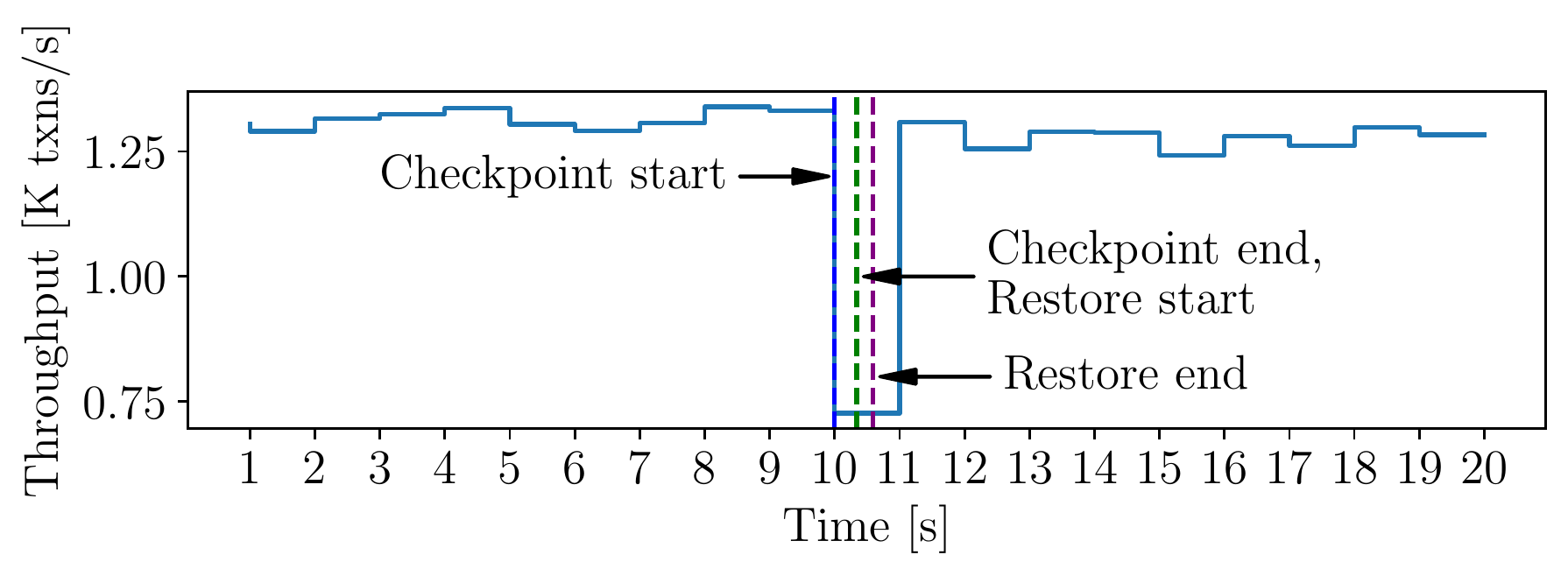}
\fi
    \caption{The blue plot shows the actual number of transactions that occurred per second. Vertical lines show the point of time where certain migration operations occur.}
    \label{fig:throughput_eval_hist}
    \end{subfigure}

    \caption{Migration affecting throughput of application with a 64 MB enclave running SmallBank benchmark.}
    \label{fig:throughput_eval_combined}
\end{figure}

We first explored how the throughput of an application is affected when it is migrated.
We created an in-memory SQLite database inside a 64 MB size enclave and manipulated the data within the database by executing the SmallBank benchmark~\cite{Alomari2008}.

Figure~\ref{fig:throughput_eval} shows the result.
The blue plot shows the number of transactions executed per second on a one second rolling average, which calculates the average of transactions in the previous second.
We plot our results starting after the first second and advance in 1 millisecond increments.
The vertical blue, green, and purple dashed line shows the time points where the \checkpoint starts, \checkpoint ends and restore starts, and restore ends, respectively.
We can see that the throughput of nearly $1,300$ transactions per second (txns/s) dips to $750$~txns/s as the migration happens, and recovers after around $1,400$ ms.
In this benchmark, when we consider the number of transaction executed per second, the rolling average of executed SmallBank transactions never dips to zero.
This shows that most applications -- which typically consider their throughput at a per-second granularity -- will never experience an instance when their application is unavailable.

We further consider a more course-gained concept of availability in Figure~\ref{fig:throughput_eval_hist}.
This figure shows the case where we plot the sum of the number of transactions recorded per second.
We see a similar trend as Figure~\ref{fig:throughput_eval}, where the number of executed transactions in a window is approximately $1,300$~txns/s, and dips to appropriately $750$~txns/s, but recovers back the next second.
Thus, we observe that a window of unavailability -- where throughput is $0$~txns/s -- does not exist and the window of reduced throughput is only one second.

\begin{figure}[t]
    \centering
\ifdefined \showFullFigure
    \includegraphics[width=\columnwidth]{figs/host_latency_with_crypto_openssl_full.pdf}
\else
    \includegraphics[width=\columnwidth]{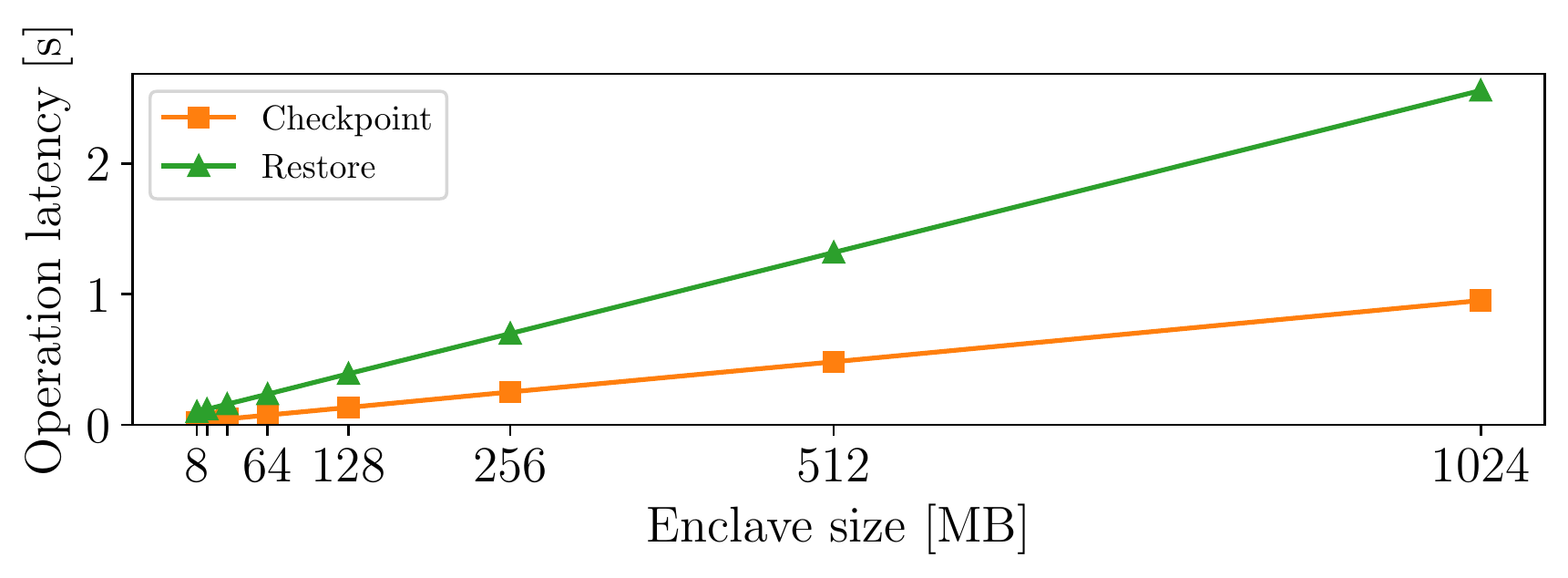}
\fi
    \caption{Latency of \checkpoint{ing}/restoring host application and enclave when varying enclave size. Enclave runs a simple counter application.}
    \label{fig:overall_latency}
\end{figure}

\begin{figure}[t]
    \centering
\ifdefined \showFullFigure
    \includegraphics[width=\columnwidth]{figs/parasite_restorer_latency_with_crypto_openssl_full.pdf}
\else
    \includegraphics[width=\columnwidth]{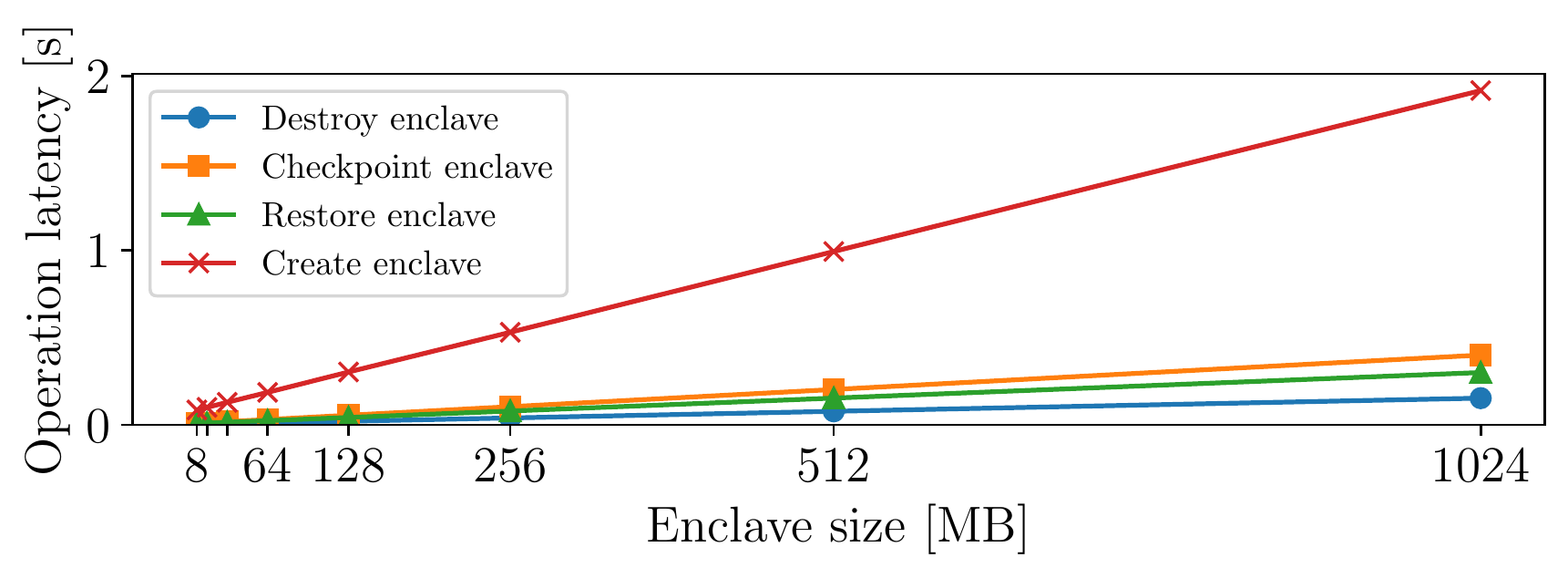}
\fi
    \caption{Latency of enclave migration operations when varying enclave size. OpenSSL is used to encrypt/decrypt enclave data. \Checkpoint and Destroy enclave operations occur during \checkpoint phase, while Create and Restore enclave operations occur during restore.}
    \label{fig:parasite_restorer_overhead_openssl}
\end{figure}

\begin{figure}[t]
    \centering
\ifdefined \showFullFigure
    \includegraphics[width=\columnwidth]{figs/host_latency_overhead_full.pdf}
\else
    \includegraphics[width=\columnwidth]{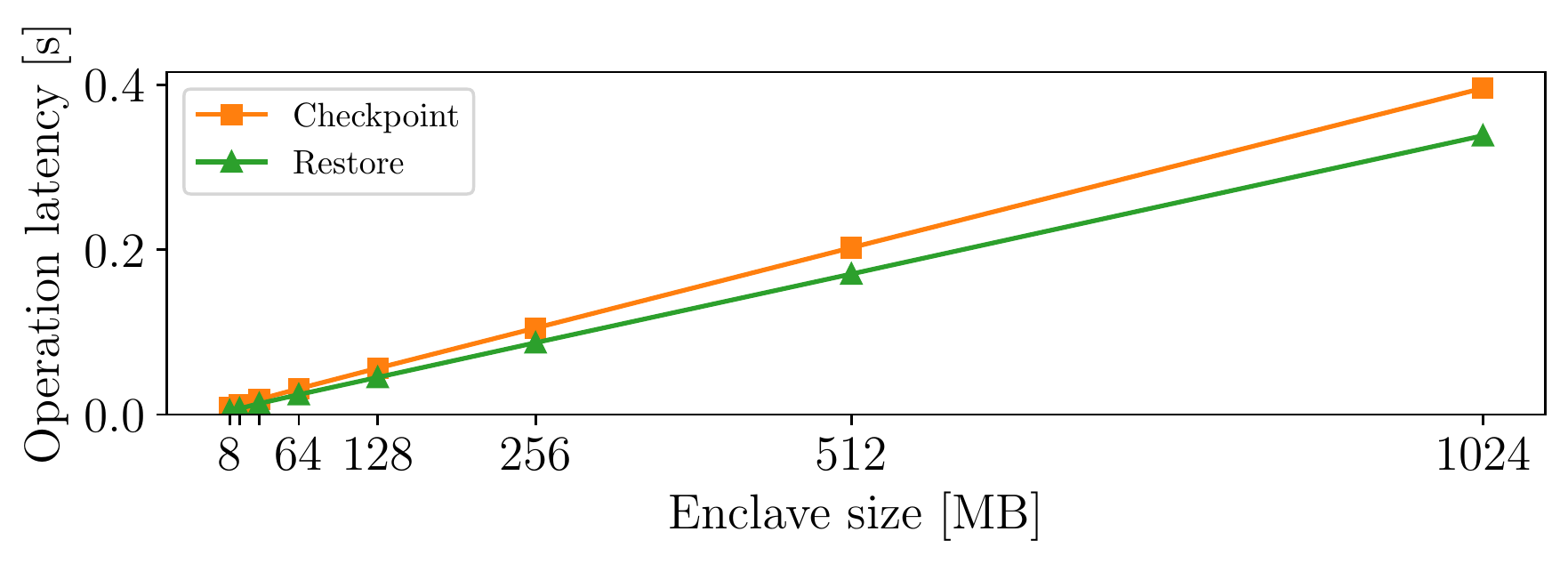}
\fi
    \caption{Latency of \checkpoint{ing}/restoring only the host application. The same amount of memory needed to store encrypted enclave data is allocated by host application and is migrated with the application.}
    \label{fig:criu_overhead}
\end{figure}

\begin{figure}[t]
    \centering
\ifdefined \showFullFigure
    \includegraphics[width=\columnwidth]{figs/host_latency_criu_no_pages_full.pdf}
\else
    \includegraphics[width=\columnwidth]{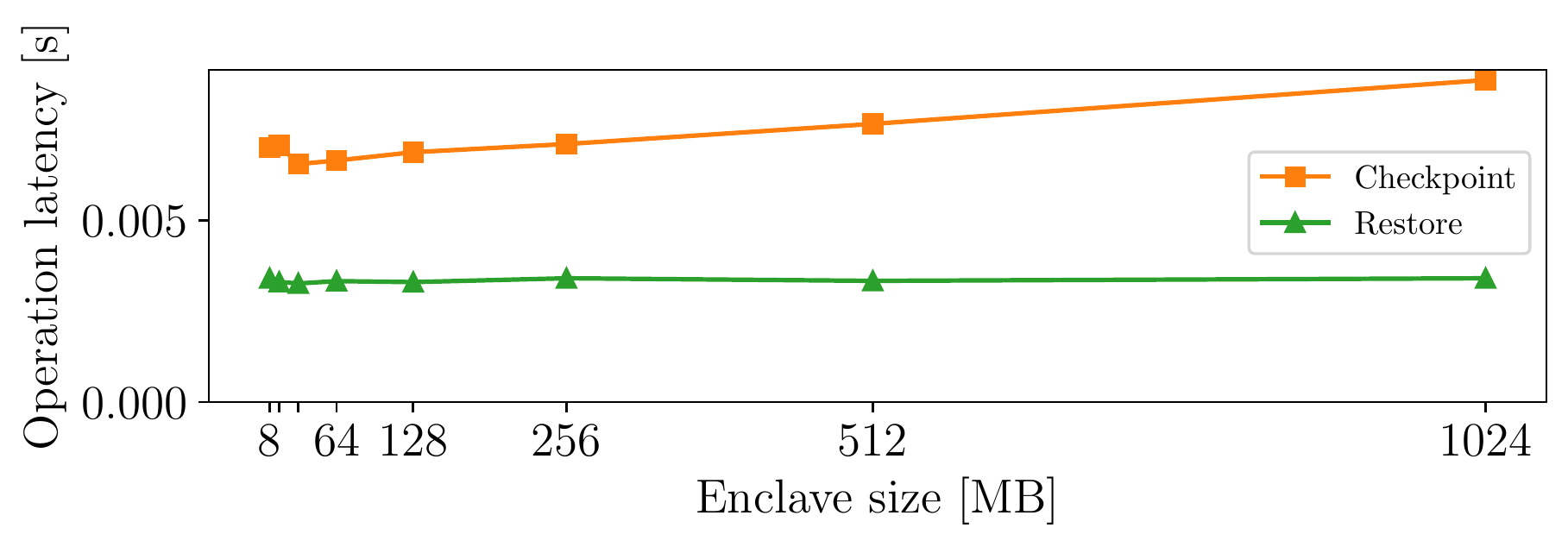}
\fi
    \caption{Latency of \checkpoint{ing}/restoring only the host application. No memory buffer is allocated.}
    \label{fig:latency_overhead_no_pages}
\end{figure}

\subsection{Migration Latency} \label{sec:latency_eval}
\label{sec:latency_vs_enclave_size}
\label{sec:overall_latency}
Now, we look the impact of migrating enclaves with an increasing amount of memory allocated to the enclave and report all results as the average of 10 measurements.
We first show the overall latency of migrating both the host application and enclave.

We measured the change in migration latency when varying the enclave size from 8 MB up to 1024 MB, increasing in two folds.
The enclave sizes were chosen based on a survey conducted by Guerreiro \etal~\cite{Guerreiro2020} on various projects that include enclaves, which they concluded that enclave sizes between 1 MB and 1 GB represent many development scenarios.
The enclave ran an application that incremented a counter from $0$ to $1,000,000,000$.
For this evaluation, we used OpenSSL (see \S~\ref{sec:latency_vs_crypto_lib} for a comparison of cryptographic libraries).
We use this benchmark to understand the overall latency of \syssgx with the side-effects that a more complex workload such as SmallBank may introduce.

Figure~\ref{fig:overall_latency} shows the time required for \syssgx to \checkpoint or \restore the host application and the enclave when the enclave size changes.
We can see that the \restore phase takes the most time ($2.55$ sec for a 1 GB enclave) while \checkpoint takes less time ($0.95$ sec).
We also observed that the latency for both phases increases linearly with the enclave size.
Since modern Icelake CPUs have an EPC which is half the available system memory, we expect this linear relationship between the enclave's memory and migration time to be consistent until memory is exhausted.
This allows \syssgx users to estimate the time required to migrate their enclave.

\subsection{Enclave Migration Latency} \label{sec:enclave_migration_latency}
Next, we consider the overhead introduced when migrating just the enclave when varying enclave size.
We measure key enclave operations, namely \checkpoint enclave, destroy enclave, create enclave, and \restore enclave.
Note that (1) and (2) occur during the \checkpoint phase and (3) and (4) occur during the restore phase.
These operations were chosen based on preliminary observations conducted prior to this evaluation.

The results are shown in Figure~\ref{fig:parasite_restorer_overhead_openssl}.
We can observe that creating an enclave takes the most time ($1.91$ sec for 1 GB enclave), followed by \checkpoint ($0.40$ sec), restore ($0.30$ sec), and destroy enclave ($0.15$ sec).
The reason that creating an enclave takes the longest is because of the allocation and initialization of enclave pages.
We can clearly see that the \restore phase is the largest contributor to the latency we observed in Figure~\ref{fig:overall_latency}.
The other essential enclave migration operations (\checkpoint and \restore enclave) introduce minimal latency.

Another point of consideration is that the create enclave latency is included in the overall latency strictly because the destination enclave is created during the migration operation.
We emphasize that this would change according to migration policies, e.g., if the policy requires the destination enclave to be created before or during migration, this latency would not occur in the critical path.
We could further reduce the latency in Figure~\ref{fig:parasite_restorer_overhead_openssl} by employing multiple threads when \checkpoint{ing}/restoring enclave data.
Therefore, the numbers reported here represent the \emph{upper bound}, where all migration operations are done sequentially.

\subsection{Host Migration Latency} \label{sec:host_migration_latency}
Next, we look at the overheads introduced when migrating only the host application. 
In this evaluation, we migrated only the host application and we did not create an enclave in the application.
Since the host application allocates and initializes a buffer required to store encrypted enclave data in \syssgx, we also wanted to observe whether this has an impact on latency.
Therefore, although the application does not create an enclave during this evaluation, the application still allocated and initialized the buffer.

Figure~\ref{fig:criu_overhead} shows the results.
We can see that the \checkpoint phase takes the longest ($0.4$ sec for a 1~GB buffer), followed by restore ($0.35$ sec).
From this we can draw several conclusions:
First, the time required to allocate and initialize the buffer required to store encrypted enclave data is significant.
This conclusion is further validated by comparing the results from Figure~\ref{fig:criu_overhead} to the results from Figure~\ref{fig:latency_overhead_no_pages} where \syssgx does not allocate the buffer.
Figure~\ref{fig:latency_overhead_no_pages} shows that latency of migrating a process without the memory buffer does not grow as the enclave's memory increases.
Second, the \checkpoint/\restore latency of a host application is near-identical to that of an enclave.
This shows \syssgx can migrate both a host application and its enclave by doubling the latency of \criu and that our implementation of in-process TEE migration has the same latency overhead as a state-of-the-art non-TEE process migration utility.

\begin{figure}[t]
    \centering
\ifdefined \showFullFigure
    \includegraphics[width=\columnwidth]{figs/parasite_restorer_latency_with_crypto_mbedtls_full.pdf}
\else
    \includegraphics[width=\columnwidth]{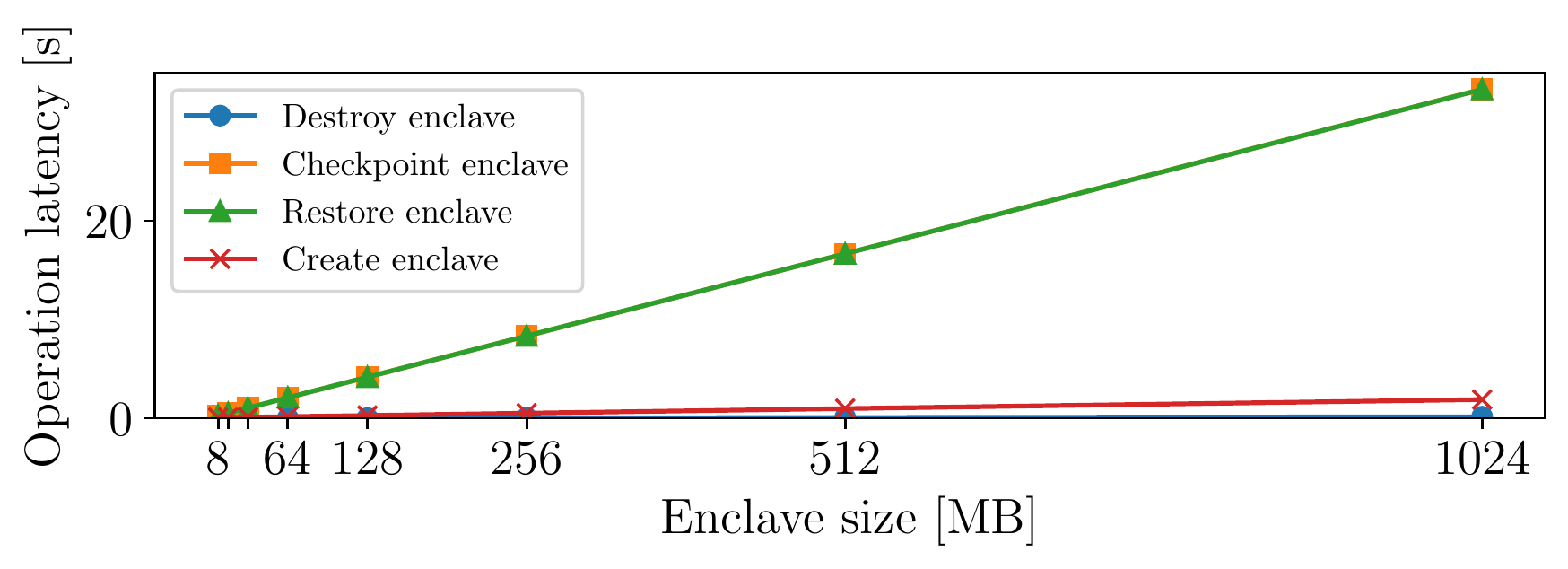}
\fi
    \caption{Latency of key enclave migration operations when varying enclave size. mbedTLS is used to encrypt/decrypt enclave data. \Checkpoint and Destroy enclave operations occur during \checkpoint phase, Create and Restore enclave operations occur during restore.}
    \label{fig:parasite_restorer_overhead_mbedtls}
\end{figure}

\begin{figure}[t]
    \centering
\ifdefined \showFullFigure
    \includegraphics[width=\columnwidth]{figs/parasite_restorer_latency_no_crypto_openssl_full.pdf}
\else
    \includegraphics[width=\columnwidth]{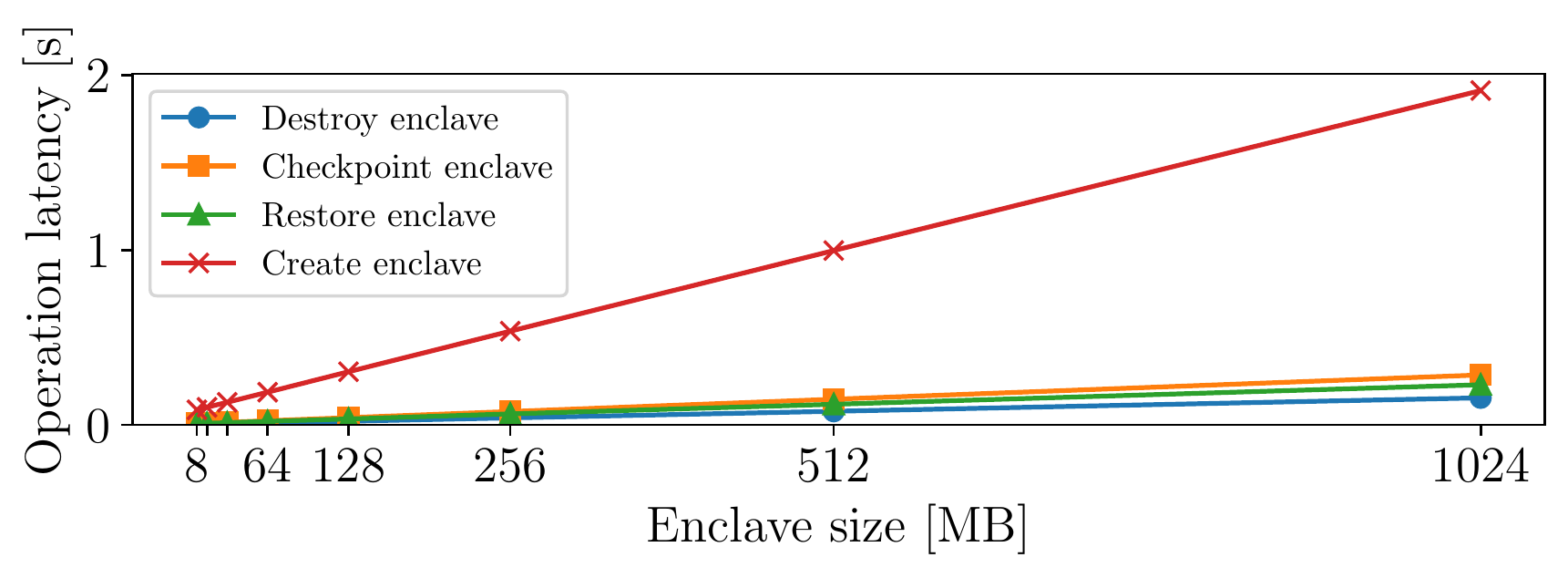}
\fi
    \caption{Latency of key enclave migration operations when varying enclave size. Enclave data is copied outside without any encryption. \Checkpoint and Destroy enclave operations occur during \checkpoint phase, Create and Restore enclave operations occur during restore.}
    \label{fig:parasite_restorer_overhead_no_crypto}
\end{figure}

\subsection{Latency vs. Crypto library} \label{sec:latency_vs_crypto_lib}
Finally, we investigate how different cryptographic libraries affect enclave migration latency.
Here we use two different cryptographic libraries supported by OpenEnclave; mbedTLS and OpenSSL.
Additionally, we measure the latency without encrypting/decrypting the enclave data during the \checkpoint/restore phase to show the overhead introduced by cryptographic operations.
The evaluation followed the same methodology as Section~\ref{sec:latency_vs_enclave_size}, where we vary the enclave size and measure the time taken to migrate an enclave.

Figures~\ref{fig:parasite_restorer_overhead_openssl} and~\ref{fig:parasite_restorer_overhead_mbedtls} show the latency of enclave migration when using OpenSSL and mbedTLS, respectively.
We see that OpenSSL outperforms mbedTLS by two orders of magnitude, as mbedTLS takes nearly $33$ seconds to \checkpoint and \restore an enclave, while it takes $0.4$ seconds to \checkpoint and $0.3$ seconds to \restore an enclave using OpenSSL.
This is validated when comparing Figures~\ref{fig:parasite_restorer_overhead_openssl} and~\ref{fig:parasite_restorer_overhead_no_crypto}, as the overhead introduced by OpenSSL is in the order of milliseconds, not seconds.

We observed from this evaluation it is preferable to use OpenSSL -- which utilizes specialized CPU instruction to perform the encryption/decryption -- whenever possible. 
MbedTLS should only be used if there are no specialized CPU instructions to accelerate cryptographic operations as mbedTLS introduces less code into the enclaves trusted code base.
This allowed us to conclude that the performance of enclave migration is dependant on the encryption and decryption performance.
Thus, we expect that, as more cryptographic operations are optimized and offloaded to hardware accelerators, the migration latency of \syssgx will decrease.

\section{Related Work} \label{sec:related}
\mypar{TEE migration}
Currently, there are only two TEE architectures (AMD SEV~\cite{SEV}, SEV-SNP~\cite{SEV-SNP}, and Intel TDX~\cite{TDX}) that have publicly announced native support for migrating their VM-based TEEs.
Park \etal and Gu \etal proposed systems that enable TEEs with Intel SGX enclaves to be migrated.
The drawback of these systems is that they either require native hardware support or TEEs to be aware and coordinate with the migration operator.
In addition, they target VM-based TEEs, not in-process ones.

TEEnder~\cite{Guerreiro2020} is the only TEE migration system which are aware of that targets in-process TEEs.
However, it uses Hardware Security Modules (HSMs) to encrypt and decrypt SGX enclave data during migration.
This hardware is not standard on commodity cloud servers and thus this reduces on which environments their solution can be used as well as its performance.

\mypar{TEE persistant state migration}
Alder et al.~\cite{Alder2017} proposed a design for including persistent state (e.g., sealed data, hardware monotonic counter values) when migrating Intel SGX enclaves.
Support for migrating such data is important, as migrated enclaves that rely on these persistent state will not be able to continue normal operation at their migration destination.
Moreover, not being able to migrate hardware monotonic counter values will undermine the security of the system, as the enclave will be susceptible against roll-back attacks.
We consider this as complementary work and envision \sys working in conjunction with this system.

ReplicaTEE~\cite{soriente2019replicatee} considers another aspect of TEE migration: provisioning of newly instantiated TEEs.
Although ReplicaTEE does not support replicating TEEs with their internal state intact, it allows server operators to start up multiple instances of the same enclave and provision them with the same secret without interacting with the application owner.
The authors have designed ReplicaTEE so that a limit is imposed on the number TEE instances that can be created.
\sys can utilize ReplicaTEE to limit the number of destination TEEs which are created while allowing such TEEs to be provisioned with necessary secrets (e.g., migration key).

\section{Conclusion} \label{sec:conclusion}
In this work, we proposed \sys, a software-only design to enable migration functionality into existing in-process TEE architectures without the need for hardware modifications.
\sys allows TEEs to be migrated at arbitrary points in their execution, allowing our method to be integrated into existing process migration tools.
We implemented \sys for Intel SGX and \criu, and show that migration latency increases linearly with the size of the TEE and is primary dependent on time required to create and initialize the destination TEE.

\ifdefined \showauthors
\fi

\footnotesize
\raggedright
\balance
\bibliographystyle{abbrv}
\bibliography{enclave_migration}

\balance

\end{document}